\documentclass [12pt,preprint]{aastex}
\usepackage{epsfig}
\usepackage{natbib}

\begin{document}
\voffset-0.5cm
\newcommand{\gsim}{\hbox{\rlap{$^>$}$_\sim$}}
\newcommand{\lsim}{\hbox{\rlap{$^<$}$_\sim$}}

\title{Long Gamma Ray Bursts Trace The Star Formation History}

 \author{Shlomo Dado\altaffilmark{1} and Arnon Dar\altaffilmark{1}}

\altaffiltext{1}{Physics Department, Technion, Haifa 32000, Israel}

\begin{abstract}

We show that if the broad-line supernova explosions of type Ic (SNeIc) 
produce the bulk of the observed long duration gamma ray bursts (LGRBs), 
which include both the high and low-luminosity LGRBs and the X-ray flashes 
(XRFs), and if LGRBs have the geometry assumed in the cannonball (CB) 
model of LGRBs, then their rate measured by Swift and their redshift 
distribution are consistent with the star formation rate (SFR) over the 
entire range of redshifts where the SFR has been measured with sufficient 
accuracy.

\end{abstract}


\maketitle

\section{Introduction} 
Core collapse supernovae are produced by the explosive death of
short-lived massive stars.  Although very bright in optical light,
ordinary core collapse supernovae are not bright enough to be resolved in
galaxies with redshift $z>1$. As such, they can be used to trace the star
formation history only up to redshifts $z\sim 1$.

Gamma-ray bursts (GRBs), the most luminous known electromagnetic events 
since the Big Bang, can be detected in MeV $\gamma$-rays up to very large 
redshifts, $z\gg 10$, with current instruments aboard satellites. Mounting 
photometric and spectroscopic evidence indicates that both long duration 
gamma ray bursts (LGRBs) and low luminosity X-ray flashes (XRFs) are 
produced by highly relativistic jets ejected in core collapse supernovae 
explosions of type Ic (SNeIc) of very massive stars at the end of their 
short life\footnote{The only known exceptions, GRBs 060614 and 060505 
(Fynbo et al.~2006; Della Valle et al.~2006; Gal-Yam et al.~2006) could 
have been produced in the death of massive stars, which generate very 
faint or failed supernovae (see, e.g., Dado et al.~2008).}. This suggests 
that the cosmic rate of LGRBs may trace the cosmic star formation rate 
(Wijers et al.~1998) back to very large redshifts beyond those accessible 
to optical measurements. However, it has been claimed that the observed 
rate of LGRBs and XRFs follows neither the rate of SNeIc nor the star 
formation rate (SFR): Unlike the SFR (in a comoving unit volume) that 
first increases with redshift, the observed LGRB rate (LGRBR) in the range 
$z < 0.1$ first decreases with increasing redshift (e.g., Guetta \& Della 
Valle~2007), while at larger redshifts it increases faster than the SFR 
(Daigne et al.~2006; Le \& Dermer~2007; Yuksel et al.~2007; Salvaterra \& 
Chincarini~2007; Li~2008; Kistler et al.~2008; Salvaterra et al.~2009). 
The discrepancy at small $z$ was interpreted as evidence that low 
luminosity LGRBs and XRFs and ordinary LGRBs with much higher luminosity 
belong to physically distinct classes (Soderberg et al.~2004; Cobb et 
al.~2006; Liang et al.~2007; Soderberg et al.~2006, Pian et al.~2006, 
Amati et al.~2007, Chapman et al.~2007). The relative rate 
$\Psi(z)$=LGRBR/SFR was claimed to behave like $\Psi(z)\propto 
(1+z)^\alpha$ with $\alpha\approx 0.5$ in the range where both are well 
observed (e.g., Robertson \& Ellis~2012, Wei et al.~2013).

The photometric evidence (see, e.g., Dado et al.~2002, Zeh et al.~2004 and 
references therein) and spectroscopic evidence (see, e.g., Stanek et 
al.~2003, Hjorth et al.~2003, Wei et al.~2013, Cenko et al.~2013) that 
broad-line SNeIc produce both low luminosity XRFs and high luminosity 
LGRBs suggests that XRFs are ordinary LGRBs viewed far off-axis (Dado et 
al.~2004), and that perhaps the different behaviour of the LGRBR and SFR 
as function of redshift, both at very low and very high redshifts, is 
because of observational selection effects. Indeed, the inferred evidence 
of a different behaviour of the LGRBR than that of the SFR at both low and 
high redshifts involved the assumption that the beaming 
fraction $f_b=<(1-cos\theta_j)>$ of detectable LGRBs is independent of 
$z$. This assumption is valid in the standard conical fireball (FB)
models of GRBs because $\theta_j$, the opening angle of the conical jet, which 
produces the observed GRB, is much larger than the relativistic beaming 
angle $\theta_b=1/\gamma$ associated with the jet bulk motion Lorentz 
factor $\gamma$. Such a jet produces an afterglow with an achromatic 
temporal break at a time that is correlated to the prompt $\gamma$-ray 
emission properties, as well as closure relations between the temporal and 
spectral behaviours before the break and after the break 
(Rhoads~1997,1999, Sari et al.~1999). But, all these predictions turned 
out to be at odds with the observed properties of the afterglows of most 
LGRBs (see, e.g., Dado \& Dar~2013 and references therein).

In the cannonball (CB) model of GRBs (Dar and De R\`ujula 2004 and 
references therein), because of relativistic beaming, Doppler shift
and time aberration, the energy-flux $F$ from a highly relativistic 
cannonball (plasmoid) with 
bulk motion Lorentz factor $\gamma\gg 1$ and Doppler factor $\delta$ 
decreases very rapidly, like $F\propto \gamma^2\, \delta^4\rightarrow 
\theta^{-8}$, when the viewing angle $\theta$ relative to the CB's 
direction of motion satisfies  $1/\gamma^2\ll\theta^2\ll 1 $. This is 
because $\delta=[1/[\gamma\,(1-\beta\,cos\theta)\approx 
2\gamma/(1+\gamma^2\,\theta^2)\propto \theta^{-2}$ for $\theta^2\ll 1$ and 
$\gamma^2\,\theta^2\gg 1$. As shown in section 3, because of this rapid 
decline of the $\gamma$-ray flux with viewing angle and the detector flux 
threshold, only a fraction $f_b(z)=1-cos\theta_{\rm max}\propto 
[D_L(z)]^{-1/2}$ of GRBs at redshift $z$  
with a luminosity above the detector threshold can be detected.  In this 
paper  we show that such a decline in the beaming fraction $f_b(z)$ of 
bi-polar LGRBs and the observed SFR that was compiled and standardized by 
Hopkins and Beacom~(2006) and by Reddy and Steidel (2009), reproduce quite 
well the observed LGRBR without invoking a relative evolution.

\section{The cosmic rate of LGRBs}
In the CB model, LGRBs are produced in SNeIc by highly relativistic 
bipolar jets of plasmoids (CBs) ejected in accretion episodes of fall-back 
material on the newly formed compact object (e.g., Dar \& De R\'ujula 2004 
and references therein). Because of relativistic beaming, such LGRBs can 
be observed only from directions near the jet direction. In SNeIc that 
produce observable GRBs, the interaction of the highly relativistic jet 
with the sub-relativistic supernova ejecta results in much higher observed 
velocities of the ejecta towards the observer and consequently in 
absorption lines of the ejecta much broader than those usually observed 
during the photospheric phase of ordinary SNeIc that are not accompanied 
by an observable GRB. If most SNeIc produce GRBs that point away from 
Earth, as assumed in the CB model, then the observed cosmic rate of GRBs 
is the cosmic rate of SNeIc (SNIcR) multiplied by the beaming factor 
\begin{equation}
{d\dot{N}_{GRB}\over dz}\approx C\,f_b(z)\, SNIcR(0)\, 
{SFR(z)\over SFR(0)}\, {dV_c(z)\over dz}\,{1\over 1+z}\,,
\label{dNdz}
\end{equation}
where $C$ is the fraction  of the sky covered by the GRB detector
and $dV_c(z)/dz$ is the comoving volume at redshift $z$.
In a standard $\Lambda$CDM cosmology, 
\begin{equation}
{dV_c(z)\over dz}={c\over H_0}\, {4\,\pi\,[D_c(z)]^2 \over 
                   \sqrt{(1+z)^3\Omega_M +\Omega_\Lambda}}\,,
\label{dVdz}
\end{equation}
where $H_0$ is the current Hubble constant, $\Omega_M$ and 
$\Omega_\Lambda$ are, respectively, the current density of ordinary energy 
and of dark energy, in critical energy-density units, and $D_c(z)$ is
the comoving distance at redshift $z$, which satisfies
\begin{equation}
D_c(z)= {c\over H_0}\, \int_0^z {dz'\over \sqrt{(1+z')^3\Omega_M 
+\Omega_\Lambda}}\,.
\label{dc}
\end{equation}

\section{The beaming factor in the CB model}
The CB model, its predictions and their extensive comparisons with 
GRB observations have been described in great detail in many 
publications (see, e.g., Dar and De R\'ujula~2004 for a review, 
Dado et al.~2009a,b for comparisons with observations).
For readers unfamiliar yet with the CB model, a short 
description of the model is enclosed in Appendix I.

In the CB model, the Doppler shift, relativistic beaming and
time aberration of the observed radiation 
yield strong dependence of the observed properties of GRBs on the bulk 
motion Lorentz and Doppler factors of the CBs (e.g., Dar and 
De R\'ujula~2000,~2004, Dado and Dar~2013 and references therein). In 
particular, 
the peak luminosity of LGRBs  satisfies $L_p= 
L_0\,\gamma^2\, \delta^4$.
Because of the detector energy-flux threshold $F_{thr}$,  
LGRBs at redshift $z$ are detectable only when 
\begin{equation} 
L_{p}= L_0\,\gamma^2\, \delta^4> 4\,\pi\,[D_L(z)]^2\,F_{thr}, 
\label{lpthr}
\end{equation}   
where $D_L(z)=(1+z)\, D_c$ is the luminosity distance to redshift $z$.
Fig.~1 presents the distribution of $L_p(z)$ of Swift LGRBs
as a function of  redshift for all Swift LGRBs, which were 
detected before November 15, 2013 and are listed
in the Greiner catalog of GRBs (http://www.mpe.mpg.de/$\sim$jcg/grbgen.html)
and whose  $L_p$ was measured by
Konus-Wind or Fermi GBM and reported in the GCN archive
(Barthelmy 1997). Also shown by lines are the lower limit behaviour
expected from Eq.~(4) and the upper limit  $max\, L_p(z)=const$
for LGRB population with no redshift evolution. 

But, $1-cos\theta\approx 1/(\gamma\, \delta)$
for  $\gamma^2\,\theta^2 \gg 1$. Consequently, Eq.~(4) implies  that
LGRBs at redshift $z$ 
are detectable only if their
viewing angle satisfies $(1-cos\theta)\leq (1-cos\theta_{max})= 
[L_0/4\,\pi\,\gamma^2\,[D_L(z)]^2\,F_{thr}]^{1/4}\propto[D_L(z)]^{-1/2}$,
which yields a beaming fraction that depends on redshift,
\begin{equation}
f_b(z)=(1-cos\theta_{max})\propto [D_L(z)]^{-1/2}\,. 
\label{bf}
\end{equation}

In the CB model, the lower limit on the Doppler factor of LGRBs $min\, 
\delta(z) \propto [D_L(z)]^{1/2}$ that follows from Eq.~(4) and yields 
Eq.~(5), also yields bounds on many other observed properties of LGRBs as 
function of redshift, which provide independent additional tests of the 
redshift dependence of the beaming fraction as given by Eq.~(5). For 
instance, in the CB model, the equivalent isotropic $\gamma$-ray 
luminosity, the peak $\gamma$-ray energy, and the break-time of the X-ray 
afterglow in the GRB rest frame satisfy, respectively, $E_{iso}\propto 
\gamma\,\delta^3$, $E'_p\propto \gamma\,\delta$ and $t'_b\propto 
1/\gamma\,\delta^2$ (e.g., Dar and De R\'ujula~2000, 2004, Dado and 
Dar~2012, and references therein). These CB model relations lead to 
correlations among these observables which were predicted long before they 
were discovered empirically. They also yield the CB model predictions, 
$min\, E_{iso}(z)\propto [D_L(z)]^{3/2}$, $min\, E'_p(z)\propto 
[D_L(z)]^{1/2}$ and $max\, t'_b(z) \propto [D_L(z)]^{-1}$, which are shown 
in Figs.~2-4 for the Swift LGRBs with known redshift that were detected 
before November 15, 2013 and are listed in the Greiner's catalog of GRBs. 
Any one of these limits can be used to extract $f_b(z)$ from the 
observations displayed in Figs.~1-4.

The normalization of $f_b(z)$, however,  can be obtained most simply from 
the CB 
model relation $E'_p\approx \gamma\,\delta \epsilon_g $ (e.g., Dar and De 
R\'ujula~2004), where $E'_p$ is the peak energy of the prompt gamma rays 
of LGRB in its rest frame (indicated by a prime), which are produced by 
inverse Compton scattering of glory photons - a light halo with a 
bremmstrahlung energy spectrum $\epsilon\, dn_\gamma/d\epsilon\propto 
exp(-\epsilon/\epsilon_g)$ surrounding the progenitor (a Wolf-Rayet star) 
that was formed by scattered stellar light from the pre-supernova 
ejecta blown from the star in eruptions sometime before its SNIc 
explosion:
If $z_m$ is the redshift of the LGRB with the lowest measured $E'_p$,  
\begin {equation}
f_b(z)\approx {\epsilon_g\over min\, E'_p}\,
\left[{D_L(z_m)\over D_L(z)}\right]^{1/2}.
\label{fbEp1}
\end{equation}
For a typical glory of Wolf-Rayet stars, 
$\epsilon_g\sim 3 $ eV corresponding to a surface temperature 
of $\sim 35,000\, K$.  XRF060218 (Campana et al.~2006)
at redshift $z=0.0331$ (Mirabal and Halpern~2006)  
had $E'_p=4.5$ keV  
the lowest $E'_p$ value measured for a  Swift GRB.
In terms of these values, Eq.~(5) can be written as 
\begin{equation}
f_b(z)=1-cos\theta_{max}\approx  6.6\times 10^{-4}\,\
    \left[{D_L(z)\over D_L(0.0331)}\right]^{-1/2}\, .
\label{fbEp2}
\end{equation}
This beaming factor is 
shown in Fig.~5 as an upper bound line
to the $z$-distribution of the $1-cos\theta$ values
extracted  from the CB model relation
$1-cos\theta\approx 1/(\gamma\, \delta)=\epsilon_g/ E'_p$
for the 134 Swift LGRBs and XRFs that were
detected before November 15, 2013, have a known redshift
listed in the Greiner catalog of GRBs
and an  $E'_p$ value measured with Konus-Wind
and/or with Fermi-GBM  and reported in the archive of GCN circulars
(Barthelmy~1997). 

Note that for $z<0.1$, where $D_L(z)\approx D_c(z)\approx c\,z/H_0 $,
Eq. (1) can be  integrated analytically over time 
and redshift to yield the approximate behaviour,  
\begin{equation}
N(<z)\approx C\,\, SNIcR(0)\,\, T\,{\epsilon_g\over 
min\,E'_{p}}\, 
{8\,\pi\over 5}\,\left({c\over H_0}\right)^3\, z_m^{1/2}\,z^{5/2}\,,
\label{Nlzllgrb}
\end{equation}
where $T$ is the total observation time. 

\section{Comparison With Observations}
{\bf Priors:} In order to compare theory and observations, we have 
adopted:\\
(a) The current best values of the cosmological 
parameters from the  Planck data (Ade et al.~2013):   
a Hubble constant $H_0=67.3\, {\rm km/s\, Mpc^{-1}} $, $\Omega_M=0.315$ 
and $\Omega_\Lambda=0.685$.\\
(b) The local rate $0.065 \pm 0.032$ SNu of SNeIc  estimated by  
Arbutina (2007), where  SNu=SN per $10^{10}\, L_\odot$ per century 
for the above value of $H_0$. For  a local luminosity density of 
$1.4\times 10^8\, L_\odot\, {\rm  Mpc^{-3}}$, it yields 
$SNIcR(0)\approx (9.0\pm 4.5)\times 10^{-6}\, {\rm Mpc^{-3}\, y^{-1}}$.\\ 
(c) The star formation rate that was compiled and standardized by Hopkins 
and Beacom~(2006) and by Reddy and Steidel (2009)
from optical  measurements.  This standardized SFR  is well 
approximated  by a  log-normal distribution, 
\begin{equation}
{\rm SFR}(z)\approx 0.25\, e^{-[ln((1+z)/3.16)]^2/0.524}\,\,
                 {\rm M_\odot\, Mpc^{-3}\, y^{-1}}\,.
\label{sfr}
\end{equation}\\    
(d) The  probability $<P(z)>\approx 262/749=0.35$
 of a Swift LGRB  to have a  measured redshift
(from emission lines of the LGRB host galaxy and from 
absorption lines or photometry of its optical afterglow):
Out of the 749 Swift LGRBs detected up to November 15, 2013,
only about 50\% have a detected optical afterglow, despite rapid optical
follow up, and only 262 have a measured redshift. Probably, 
most of the "missing redshifts" are due to dust extinction, the
limiting sensitivity of the telescopes, and the time it takes to acquire 
spectroscopic/photometric redshifts (e.g., Coward et al.~2012).
From the relative number of LGRB redshift measurements 
by the three methods as function of $z$ reported in 
GCN catalog (Barthelmy~1997), we have estimated 
that $P(z)\approx 0.684\, e^{-z/0.5}+0.316$. However, the use of this 
rough estimate instead of $P(z)=<P(z)>$ has a negligible effect 
on our results, except for $z\lsim 0.1$ (see below).  

{\bf Tests of the SFR-LGRBR connection:} Because of the 
different sensitivities and sky coverage of different GRB missions, we 
have limited our comparison to the 262 LGRBs and XRFs  with known 
$z$ that were detected by Swift before November  15, 2013
and are listed in the Greiner catalog of GRBs. For this choice:

\noindent
{\bf The daily rate of LGRBs} obtained by integrating  Eq.~(1) over 
redshift, using Eqs.~(2),(3),(7) and the priors (a)-(c), 
is LGRBR$\approx 2.1\pm 1.5\, {\rm day^{-1}}$ of observable 
LGRBs (high luminosity GRBs + low luminosity GRBs + XRFs) for a full sky 
coverage ($C=1$). The 749 LGRBs detected by Swift before November 15, 2013 
with a sky coverage of $C\approx  1.4/4\,\pi=0.11$ (Salvaterra \& 
Chincarini~2007), yield 
$\approx 2.07\, {\rm day^{-1}}$ Swift-like LGRBs over the entire sky,
in agreement with that obtained from  Eq.~(1).

\noindent
{\bf The redshift distribution of Swift LGRBs with known redshift} is 
compared in Fig.6~ to that predicted by Eqs.~(1)-(3),(7)  with the 
priors (a)-(d). The predicted mean 
redshift, $<z>=2.08$, is consistent with $<z>=2.12$ obtained from 
the reported redshifs of the 262 
Swift LGRBs with known redshift, which are listed in the Greiner GRB 
catalog. The 
agreement between the expected and the observed distribution is quite good 
($\chi^2/df=25.5/26=0.98$, significance level 49\%, assuming Poisson 
statistics, i.e., $\sigma_i=\sqrt{N_i}$ where $N_i$ is the observed number 
of LGRBs in bin i).  The bins 21-30 and 31-50 were converted to two bins 
$5<z\leq 7$ and $7<z\leq 10$, respectively, in order to have $N_i\geq 5$. 
Similar agreement ($\chi^2/df=48.88/49=1.00$, significance level 48\%) was 
obtained when all the 50 $z$ bins of width $\Delta z=0.2$ shown in Fig.~6 
were included in the $\chi^2/df$ calculation but the standard deviation 
errors $\sigma_i=\sqrt{N_i}$ were estimated from the predicted value of 
$N_i$. Similar agreement ($\chi^2/df=7.56/7=1.08$, with a significance 
level of 37\%, and $\chi^2/df=5.02/5=1.00$ with a significance level of 
41\%, respectively) were obtained for the TOUGH (56 Swift LGRBs, Hjorth et 
al.~2012) and BAT6 (52 Swift LGRBs, Salvaterra et al.~2012) samples when 
they were compared with the CB model predicted distribution.

Fig.~6, however, indicates  $\sim 8$ LGRBs deficiency of Swift LGRBs with 
measured redshift in the bin $1.8\leq z\leq 2.0$, in the so called 'GRB 
redshift desert', which is believed to be due to an observational bias 
rather than a real deficiency (Coward et al.~2012). Indeed, the TOUGH, and 
in particular the BAT6 sample of bright Swift LGRBs (Salvaterra et 
al.~2012), have a redshift distribution where the 'GRB desert' is almost 
completely filled. Adding eight 'missing LGRBs' to the $[1.8<z<2.0]$ bin 
of the Swift distribution of 262 LGRBs improves the good agreement 
reported above to a remarkable agreement between the differential 
distribution predicted by the CB model and the observed distribution 
($\chi^2/df=15.8/24=0.66$, significance level  90\%). 
 
\noindent 
{\bf The cumulative distributions $N(<z)$} of the 262 and 270 
(262+8 missing) Swift LGRBs with known redshift, which were detected 
before November 15, 2013, and the cumulative distributions 
predicted by the CB model (using Eqs.~(1)-(3),(7) 
with the priors (a)-(d)) are compared in Figs.~7 and 8, respectively. 
Also shown 
in Figure~8 are the expected  distribution in the standard fireball model 
with an assumed $z$-evolution of the LGRBR relative to the SFR of the form
LGRBR/SFR$\propto (1+z)^\alpha$ with $\alpha=0$ (no evolution) or 
$\alpha=1/2$ (a best fit obtained by Robertson and Ellis (2012) and by Wei 
et al.~(2013) to their selected samples of 112 and 86 bright Swift LGRBs, 
respectively, in the range $z\leq 4$. As can be seen by simple 
inspection of Figs.~7 and 8, the agreement between the observed and the CB 
model distributions of the complete sample of Swift LGRBs with known 
redshift 
($0.0331\leq z\leq 9.2$), is very good and becomes remarkable once the 
'eight missing GRBs' are added to the $[1.8<z<2.0]$ bin\footnote{ 
The sum of independent variables with a                           
Poisson distribution is also a Poisson distribution with a mean and 
a variance equal to the sum of means and variances of the independent
variables. Hence the standard deviation error of N($<z$) 
is $\sqrt{N(<z)}$. Moreover, the significance level of a     
$\chi^2$ goodness of fit to the binned differential distribution  
is also the significance level of agreement between
the corresponding binned theoretical and measured 
cumulative distribution functions (CDFs).}.

The cumulative distribution shown in Fig.~8
allows alternative formal tests of goodness 
of fit, such as the Kolmogorov-Smirnov (KS) and Anderson-Darling 
(AD) statistical tests. Such tests also yield very high significance 
levels for 
the agreement between the observed CDF and the CB model CDF and rejection 
of the FB model CDFs with $\alpha\geq 0$ relative evolution. E.g., the KS 
statistic of the CB model fit, Dmax=0.0436 for n=262 yields a significance 
level $>90\%$, while models with $f_b=const$ with $\alpha$=0, 0.5 and 1.5 
yield Dmax=0.151, 0.222, and 0.382, respectively with a significance level 
$\ll 1\%$.

\noindent
{\bf The observed rate of low-luminosity LGRBs and XRFs} was compared 
in Fig.~9 with that predicted by Eq.~(7), assuming that the probability 
of obtaining the redshift of the host galaxies of very nearby LGRBs is 
$P(z)\approx 1$ rather than the mean value $P(z)\approx 0.35$ for the 
entire $0\leq z\leq 10$ range. Also, because of the limited statistics, 
the 
true $f_b(z)$  for Swift LGRBs may lie somewhat above the upper limit 
shown in Fig.~5. Indeed, the lowest reported $E'_p$ value of LGRB 
was $E'_{p}=3.37\pm 1.79$ keV  for XRF020903 at 
$z\,=\,0.25$ measured with the HETE satellite (Amati et al.~2002). 
With $f_b$ estimated from these values and $P(z)=1$, Eq. (6)  yields 0.33, 
1.4, 3, and 3.9  expected detections ($\pm 70\%$ 
estimated error) of LGRBs+XRFs  with redshift smaller than  
0.0331, 0.059, 0.080, and 0.089, respectively, which are the redshifts of
XRF060218 (Mirabal and Halpern~2006), GRB100316D (Vergani et al.~2010), 
XRF051109B (Perley et al.~2006) , GRB060505 (Ofek et al.~2007), the 
lowest $z$ LGRBs that were  detected by Swift during 
$T\approx 9$ years of observations (see the Greiner GRB catalog, 
http://www.mpe.mpg.de/$\sim$jcg/grbgen.html). The above predicted values   
are in agreement with the corresponding values 0, 1, 2, and 3 of the 
observed cumulative distribution.  Moreover, 
since the launch of the BeppoSAX in 1996 only one GRB 
(980425 at $z=0.0085$) was detected by  the $\gamma$-ray 
satellites at a redshift smaller than 
0.0331 during a combined effective observation time of roughly
$T\approx 17$ Swift observation years, compared to 
$0.66\pm 0.46$  expected  in the CB model.

\section{Discussion and conclusions} 
Star formation at redshifts $z>6$ may have made an important contribution 
to the reionization of the universe, to its optical depth to the cosmic 
background radiations and to its metalicity at high redshifts. However, 
direct measurements of the star formation rate at redshifts $z\gsim 6$ and 
their correct interpretation are quite challenging (for a review, see, 
e.g., Robertson et al. 2010). Long duration GRBs, which are produced 
in SNeIc explosions of massive stars, and whose optical afterglows 
are visible at redshifts, which exceed by far those where direct 
measurements of the SFR are still possible, offer the possibility to 
extend the SFR 'measurements' to redshifts far beyond those of the direct 
measurements. 

The measured rate of LGRBs as a function of 
redshift, however, was claimed to differ significantly from the observed 
SFR, both at low and high redshifts. 
The discrepancy at low redshift (approximately a factor $\sim 100$)
was explained by assuming that LGRBs with 
low isotropic luminosities belong to a class different from that of LGRBs 
with high isotropic equivalent luminosities, which 
is at odds with the observations that 
both the low-luminosity and the high-luminosity LGRBs are produced by 
similar broad-line SNeIc. Moreover, the assumption does not explain why 
the observed  rate as function of redshift of low-luminosity LGRBs, which 
are produced by SNeIc, does not follow that of the SFR at $0<z<0.1$.

In the range $0.1\leq z\leq 4$ , the LGRBR was claimed to have a more 
rapid evolution relative to the SFR, which can be well parametrized 
by $(1+z)^{0.5}$. 
But, such a relative evolution (assuming a $z$-independent beaming 
factor), which was shown to fit well selected
samples of bright Swift LGRBs with a known redshift in the range 
$z<4$ (Robertson \& Ellis~2012, Wei et al.~2013) fails to describe 
the complete distribution of  Swift LGRBs with known redshift, which 
currently extends over $0.0331\leq z \leq 9.2$.
In fact, FB models with $z$-independent beaming factor and 
$\alpha\geq 0$ overpredict the high $z$  SFR 
inferred from the abundance of UV-selected galaxies
(Robertson \& Ellis~2012 and references therein).

In this paper, however, we have shown that the above discrepancies 
may have been the result of assuming a redshift-independent beaming 
factor of  LGRBs, which was adopted from the underlying current 
geometry of the standard conical fireball 
model of LGRBs. Once the assumed conical geometry of LGRBs is replaced by the 
geometry adopted in the cannonball model of GRBs, the beaming factor of 
LGRBs becomes $z$-dependent as a result of the detection threshold of 
GRBs.  We have shown that this plus the assumption that all SNeIc produce 
LGRBs, most of which are beamed away from Earth, yields a theoretical rate 
of LGRBs, which correctly reproduces: (i) the full sky rate of cosmic 
LGRBs ($\approx 2.1\, {\rm day^{-1}})$ above the Swift detection 
threshold, (ii) the observed distribution of Swift LGRBs with known 
redshift as a function of the redshift, and consequently (iii) the 
observed mean redshift $<z>\approx 2.12$ of Swift LGRBs with known $z$, 
and (iv) the cumulative distribution $N(<z)$ of Swift LGRBs as a function 
of $z$ between their lowest and highest observed redshifts.

We conclude that LGRBs seem to trace the SFR as a function of redshift at 
least up to $z\approx 6$, that probably most SNeIc produce LGRBs beamed 
away from Earth, and that the high-luminosity LGRBs and the low-luminosity 
LGRBs and XRFs seem to belong to the same class of SNIc-GRBs while the 
observed differences between them are produced by the strong dependence of 
the observed LGRB properties on detector threshold and on their Lorentz 
factor and viewing angle, which are well described by the cannonball model 
of GRBs.

\section{{\bf Appendix I} - Outline of the CB model of LGRBs}

In the cannonball (CB) model of GRBs (Dado et al.~2002; Dar and De 
R\'ujula~2004; Dado et al.~2009a,b), LGRBs and their afterglows are 
produced by the interaction of bipolar jets of highly relativistic 
($\gamma>>1$) plasmoids (CBs) of ordinary matter with the radiation and 
matter along their trajectory (Shaviv and Dar~1995, Dar~1998). Such jetted 
CBs are presumably ejected in accretion episodes on the newly formed 
compact stellar object in core-collapse supernova (SN) explosions (Dar et 
al.~1992, Dar and Plaga~1999, Dar and De R\'ujula~2000). it is 
hypothesized that an accretion disk or a torus is produced around the 
newly formed compact object, either by stellar material originally close 
to the surface of the imploding core and left behind by the 
explosion-generating outgoing shock, or by more distant stellar matter 
falling back after its passage (Dar and De R\'ujula~2000,~2004). As 
observed 
in microquasars, each time part of the accretion disk falls abruptly onto 
the compact object, two CBs made of ordinary-matter plasma are emitted in 
opposite directions along the rotation axis from where matter has already 
fallen back onto the compact object due to lack of rotational support. The 
prompt -ray pulses and early-time X-ray flares are dominated by inverse 
Compton scattering (ICS) of glory photons - a light halo surrounding the 
progenitor star that was formed by stellar light scattered from the 
pre-supernova ejecta/wind blown from the progenitor star - by the CBs 
electrons. The ICS is overtaken by synchrotron radiation (SR) when the CB 
enters the pre-supernova wind/ejecta of the progenitor star.

\noindent
{\bf Acknowledgement}: We thank an anonymous referee for useful
comments and suggestions.

{}

\newpage 
\begin{figure}[]
\centering
\vspace{-2cm}
\epsfig{file=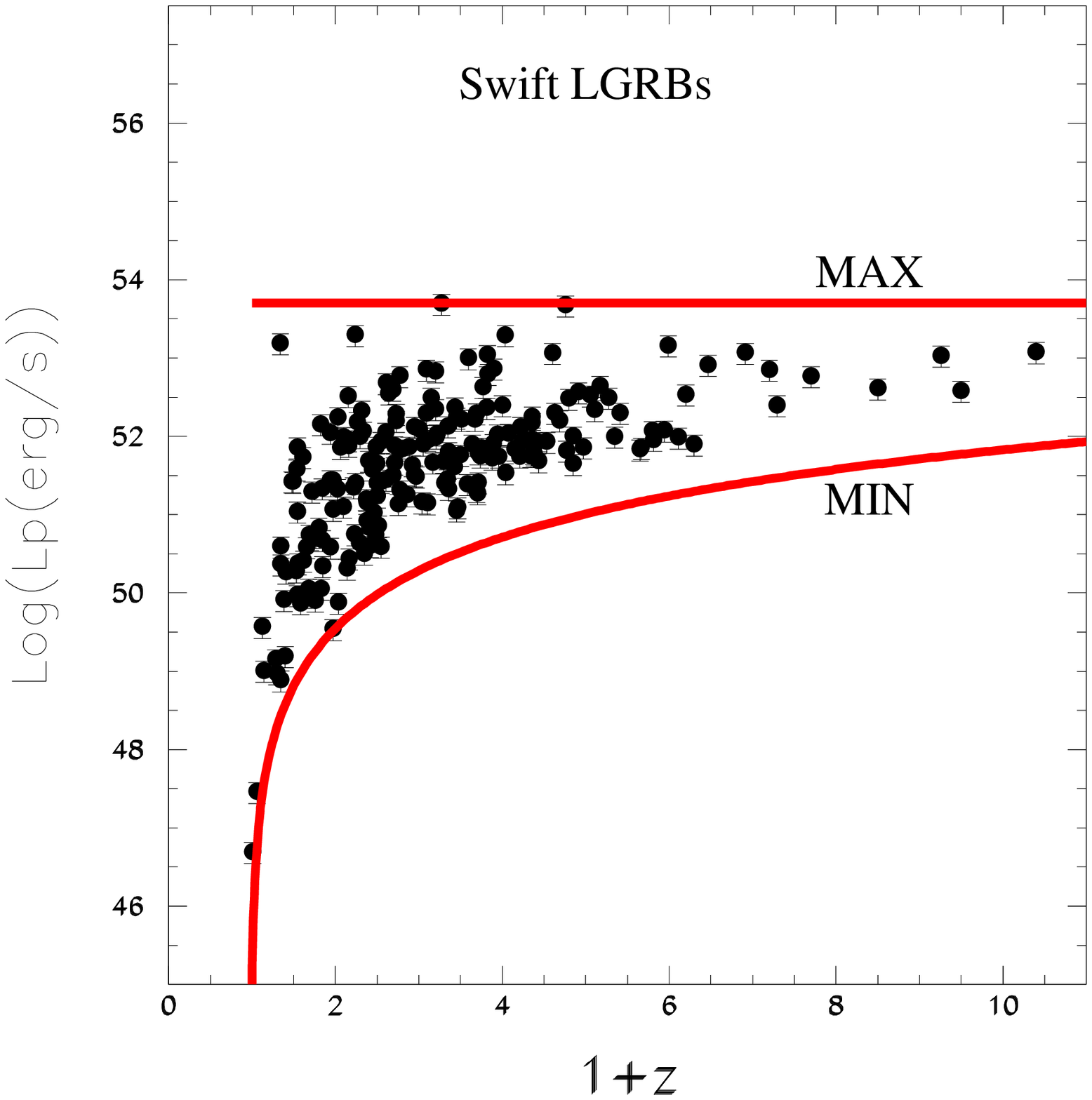,width=16.cm,height=16.cm}
\caption{
The distribution of the isotropic equivalent peak gamma-ray luminosity 
$L_p(z)$ as function of redshift of Swift LGRBs which were detected before 
November 15, 2013 and are listed in the Greiner catalog of GRBs, and whose 
$L_p$ was measured With Konus-Wind and/or Fermi GBM and reported in the 
GCN archive (Barthelmy 1997). Also shown are 
the best fit lower limit line $min\, L_p(z)\propto [D_L(z)]^2$ and the 
upper limit line $max\, L_p=const$ line expected in the CB model of 
LGRBs.}
\label{Fig1}
\end{figure}

\newpage
\begin{figure}[]
\centering
\vspace{-2cm}
\epsfig{file=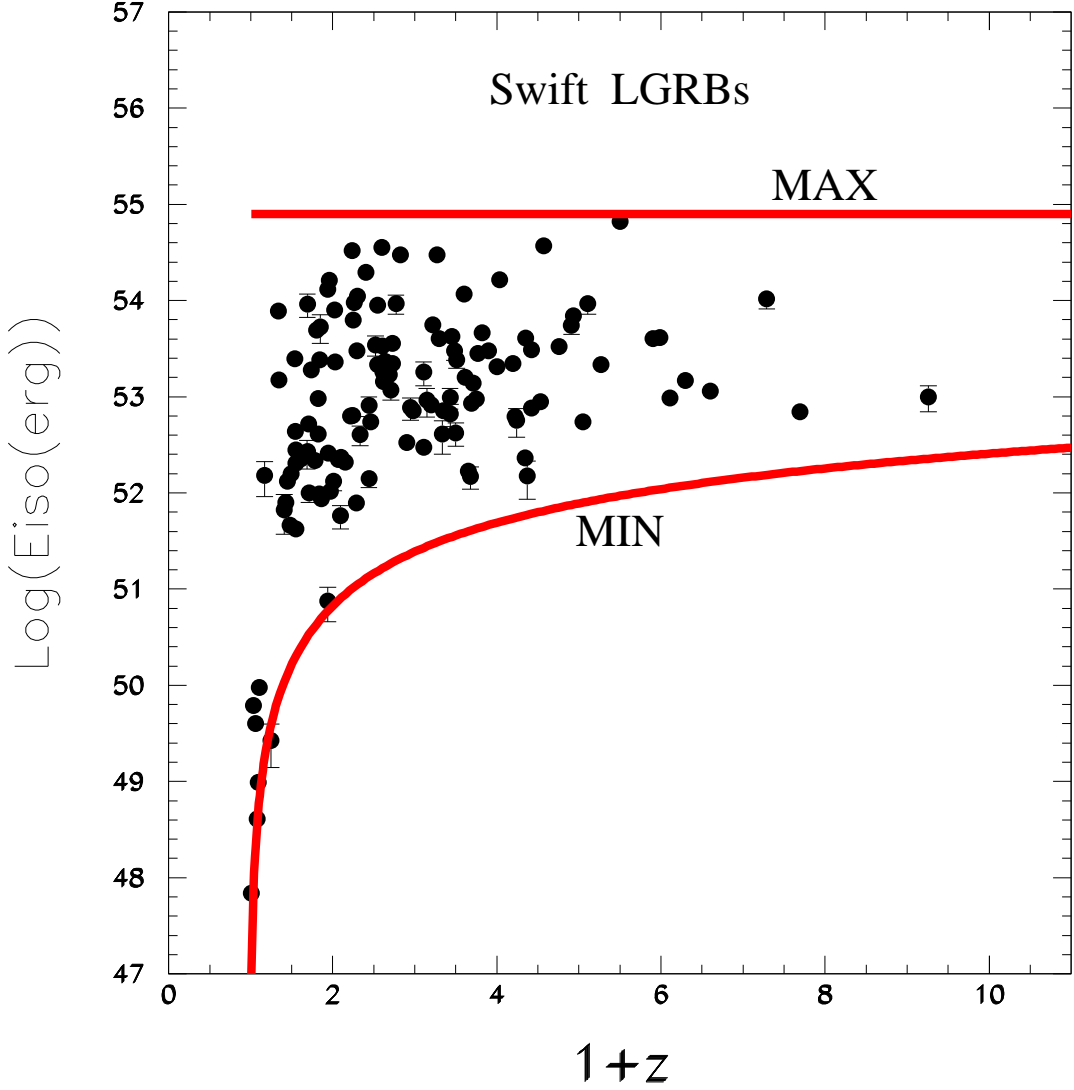,width=16.cm,height=16.cm}
\caption{
The distribution of the isotropic equivalent $\gamma$-ray energy $E_{iso}$
as a function of  redshift for all LGRBs 
which were detected
by Swift before November 15, 2013 and are listed
in the Greiner catalog of GRBs, and  whose redshift and $E_{iso}$ were
measured.  Also shown are the CB model 
best fit lower  and upper limit lines, $min\, E_{iso}(z)\propto 
[D_L(z)]^{3/2}$, and $max\, E_{iso}(z)=const$, respectively,
of  the measured $E_{iso}$ distribution  as function of $z$.}
\label{Fig2}
\end{figure}

\newpage
\begin{figure}[]
\centering
\vspace{-2cm}
\epsfig{file=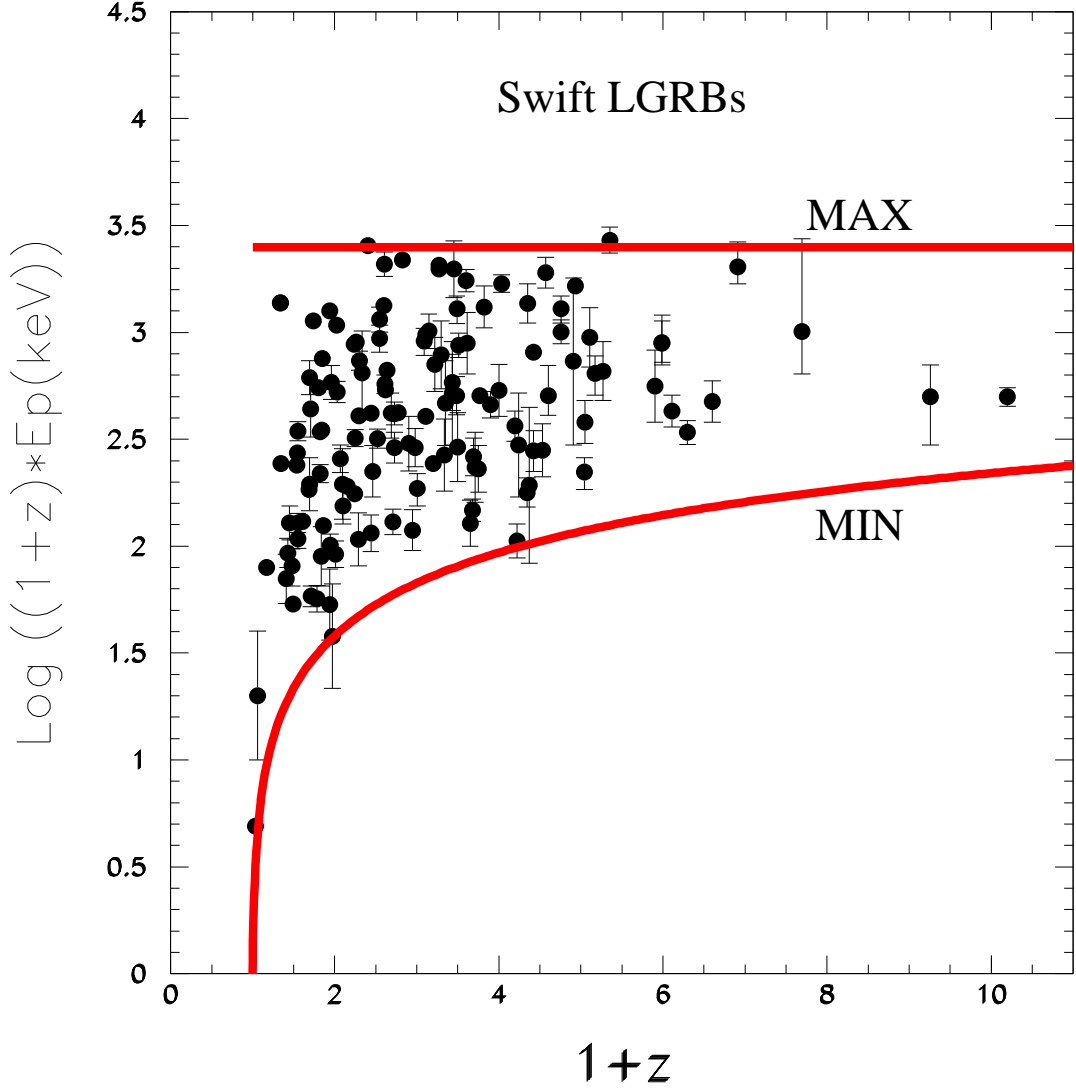,width=16.cm,height=16.cm}
\caption{    
The distribution of the $E'_p(z)$ of 134 Swift LGRBs
with measured redshift and  $E'_p$, which were detected before 
November 15, 2013
and are listed in the Greiner catalog of GRBs. The lines
represent the best fit CB model lower and upper limits, 
$min\, E'_p(z)\propto [D_L(z)]^{1/2}$ and $max\, E'_p(z)=const$,
respectively, to the measured $E'_p(z)$
distribution of Swift LGRBs as function of $z$.}
\label{Fig3}
\end{figure}

\newpage
\begin{figure}[]
\centering
\vspace{-2cm}
\epsfig{file=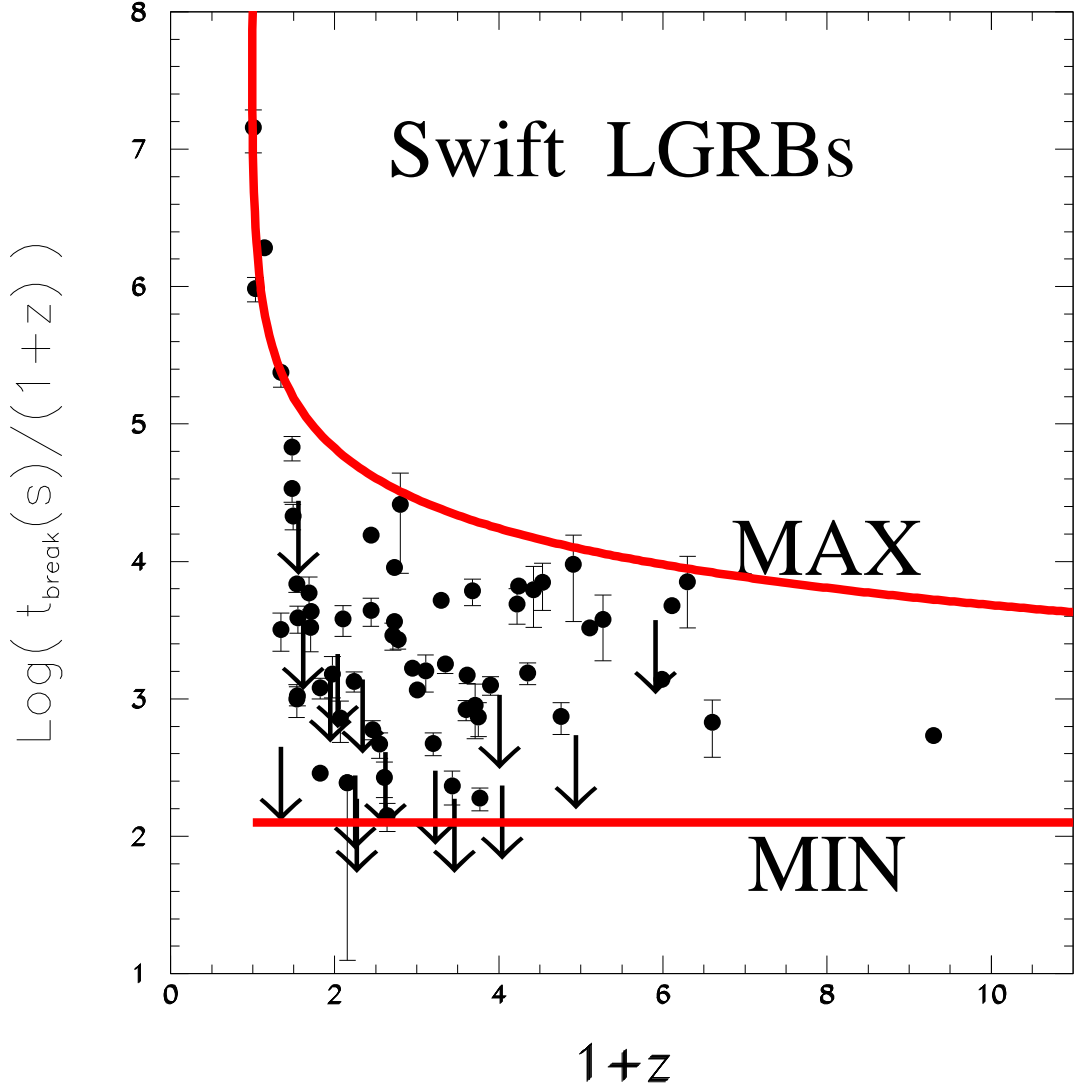,width=16.cm,height=16.cm}
\caption{
The distribution of the X-ray afterglow break time $t'_b(z)$ of 74 Swift
LGRBs  that were detected before November 15, 2013
and are listed in the Greiner catalog of GRBs. The lines
represent lower and upper limits to the
observed distribution of Swift GRBs expected in the CB model. Also
shown (the highest point) is the break time of the X-ray afterglow of 
GRB980425 
that
was measured with the Chandra X-ray telescope (Kouveliotou et al. 2004)}.
\label{Fig4}
\end{figure}

\newpage
\begin{figure}[]
\centering
\vspace{-2cm}
\epsfig{file=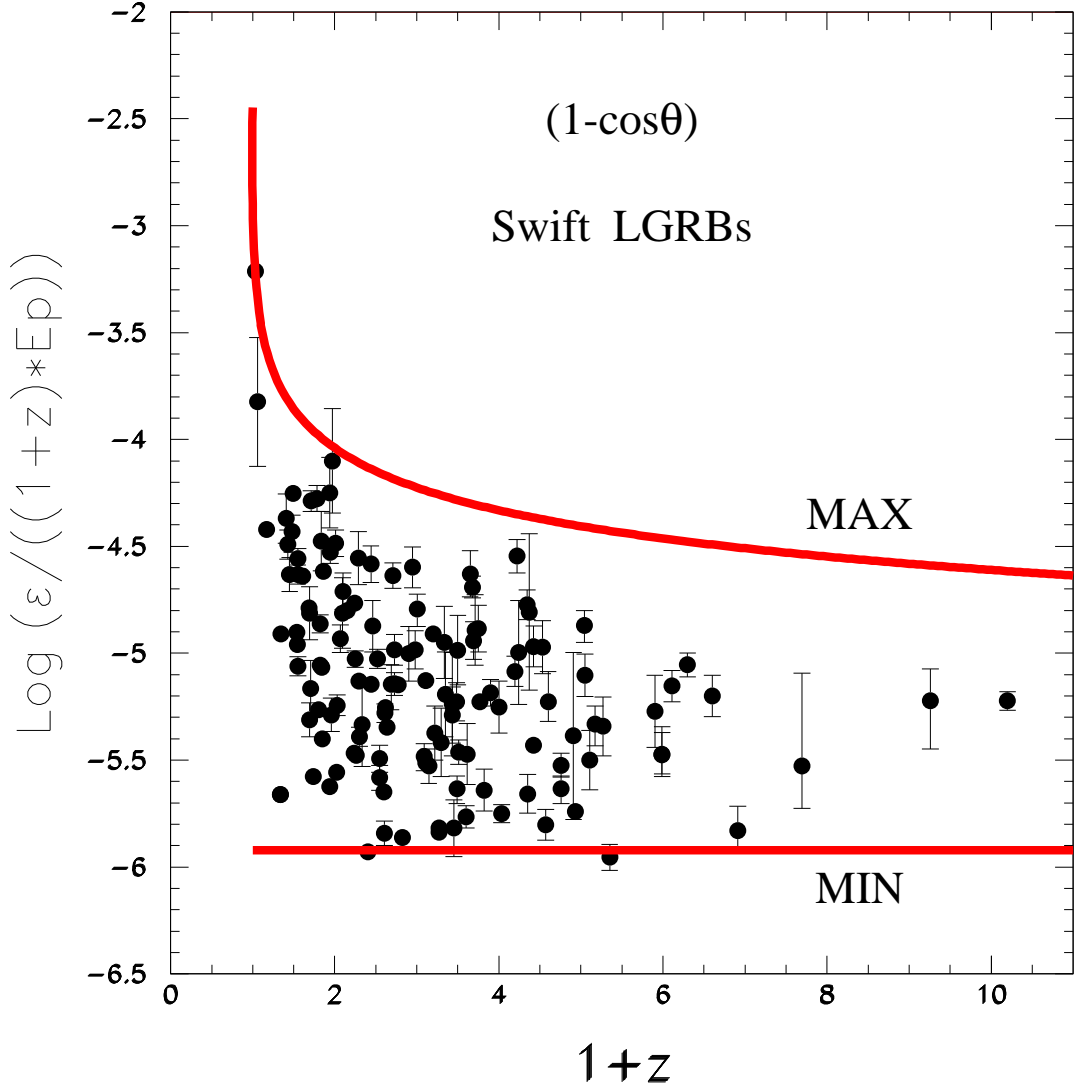,width=16.cm,height=16.cm}
\caption{
The distribution of the ratio
$\epsilon_g/E'p(z)$ for the 134 Swift LGRBs
with measured redshift and $E'_p$, which were detected before 
November 15, 2013 and are listed in the Greiner catalog of GRBs.
The lines represent  the lower and upper limits to the 
distribution expected in the CB model of GRBs.
The upper limit $MAX \propto [D_L(z)]^{-1/2}$  is the expected beaming 
factor $f_b(z)$ in the CB model.} 
\label{Fig5}
\end{figure}

\newpage
\begin{figure}[]
\centering
\vspace{-2cm}
\epsfig{file=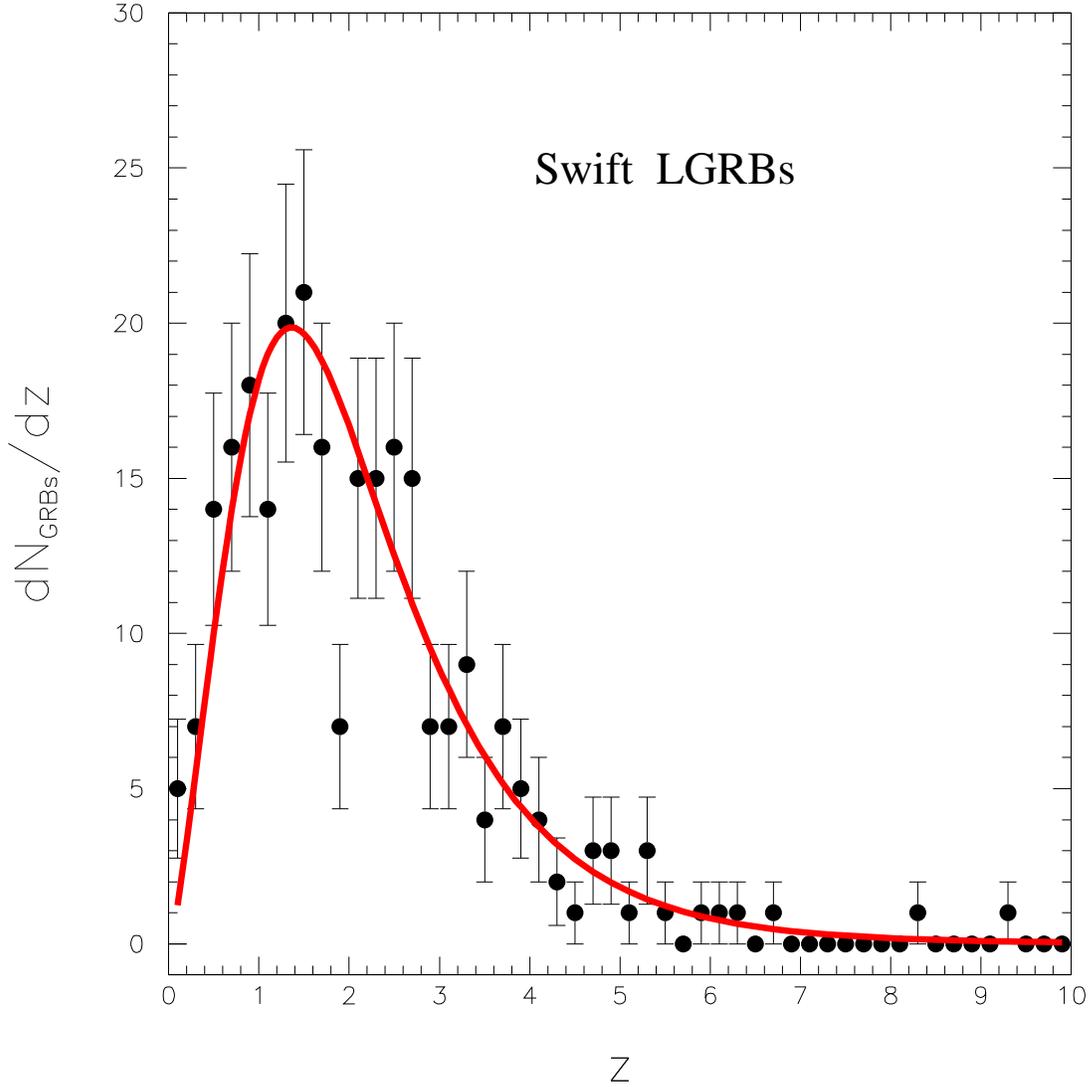,width=16.cm,height=16.cm}
\caption{The redshift distribution of the 262 Swift 
LGRBs with known redshift that were detected before 
November 15, 2013  and the 
expected distribution in the CB model of LGRBs produced in SNeIc whose 
rate traces the SFR.}
\label{Fig6}
\end{figure}

\newpage
\begin{figure}[]
\centering
\vspace{-2cm}
\epsfig{file=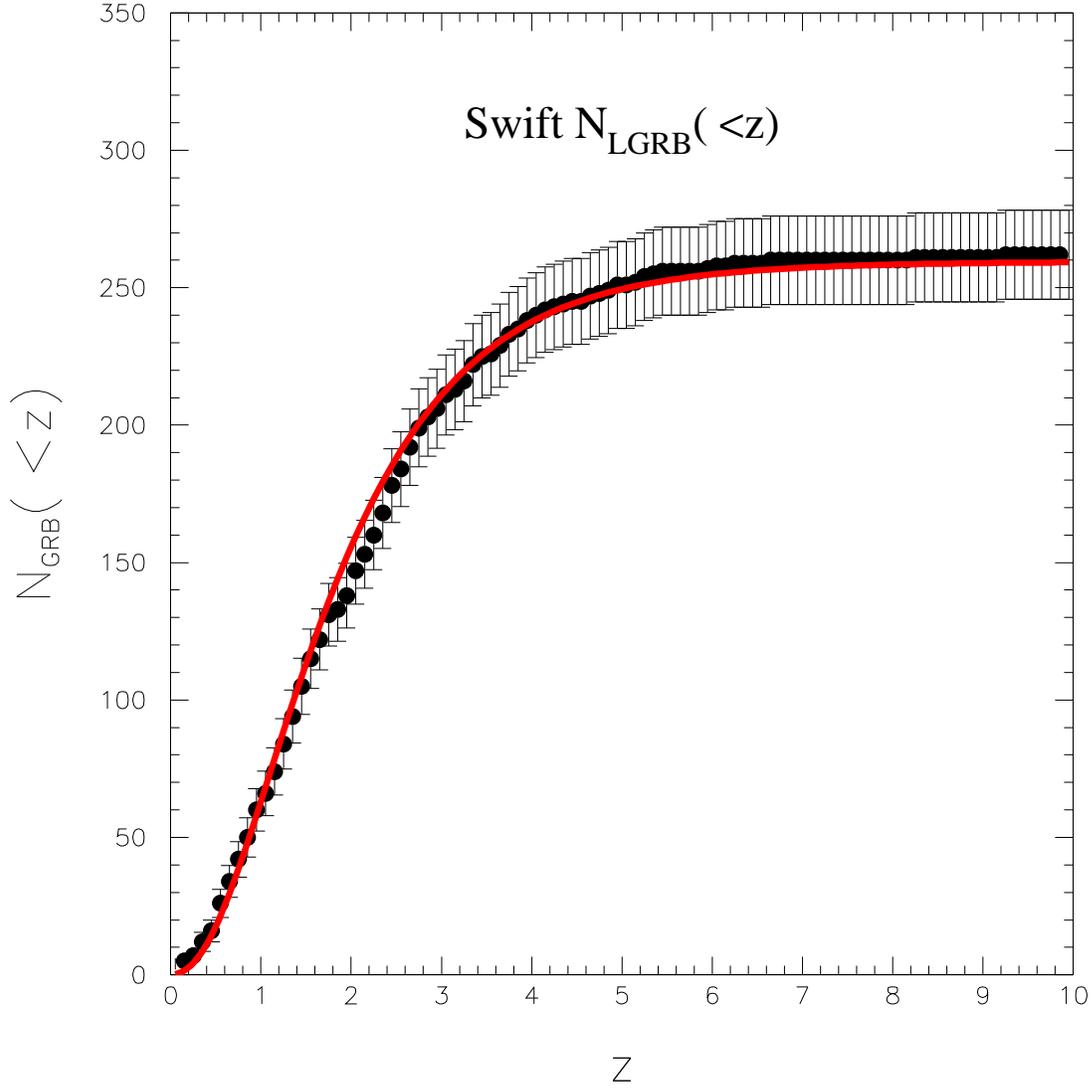,width=16.cm,height=16.cm}
\caption{
Comparison between the  
cumulative distribution $N(<z)$ of the 262 Swift LGRBs with 
known redshift that were detected  before November 15, 2013
and the distribution  
expected in the CB model of LGRBs produced in  SNeIc with a rate that 
traces the SFR.}
\label{Fig7} 
\end{figure} 

\newpage
\begin{figure}[]
\centering
\vspace{-2cm}
\epsfig{file=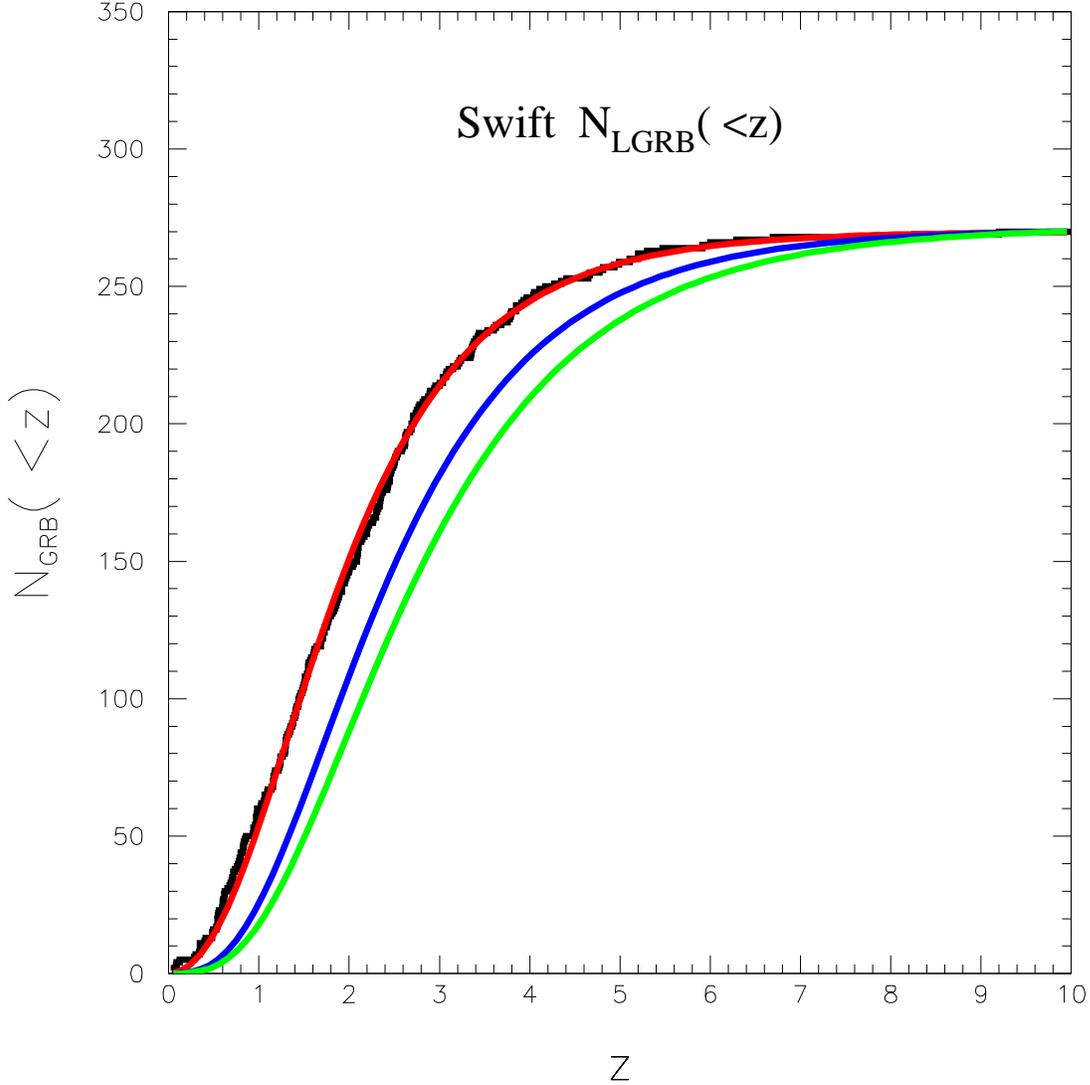,width=16.cm,height=16.cm}
\caption{Comparison between the cumulative distribution function $N(<z)$ 
of the 262 LGRBs with known redshift (histogram)
that were detected by Swift 
before November 15, 2013 after adding 8 'missing' 
LGRBs with $1.8\leq z \leq -2.0$, and the corresponding $N(<z)$  expected 
in the CB model (left curve) for LGRBs  produced in SNeIc whose rate 
traces the SFR. Also shown are the distributions expected in 
the FB model  
with a relative evolution ($\alpha=0.5$, right curve) and with no 
evolution ($\alpha=0$, middle curve).}
\label{Fig8}
\end{figure}

\newpage
\begin{figure}[]
\centering
\vspace{-2cm}
\epsfig{file=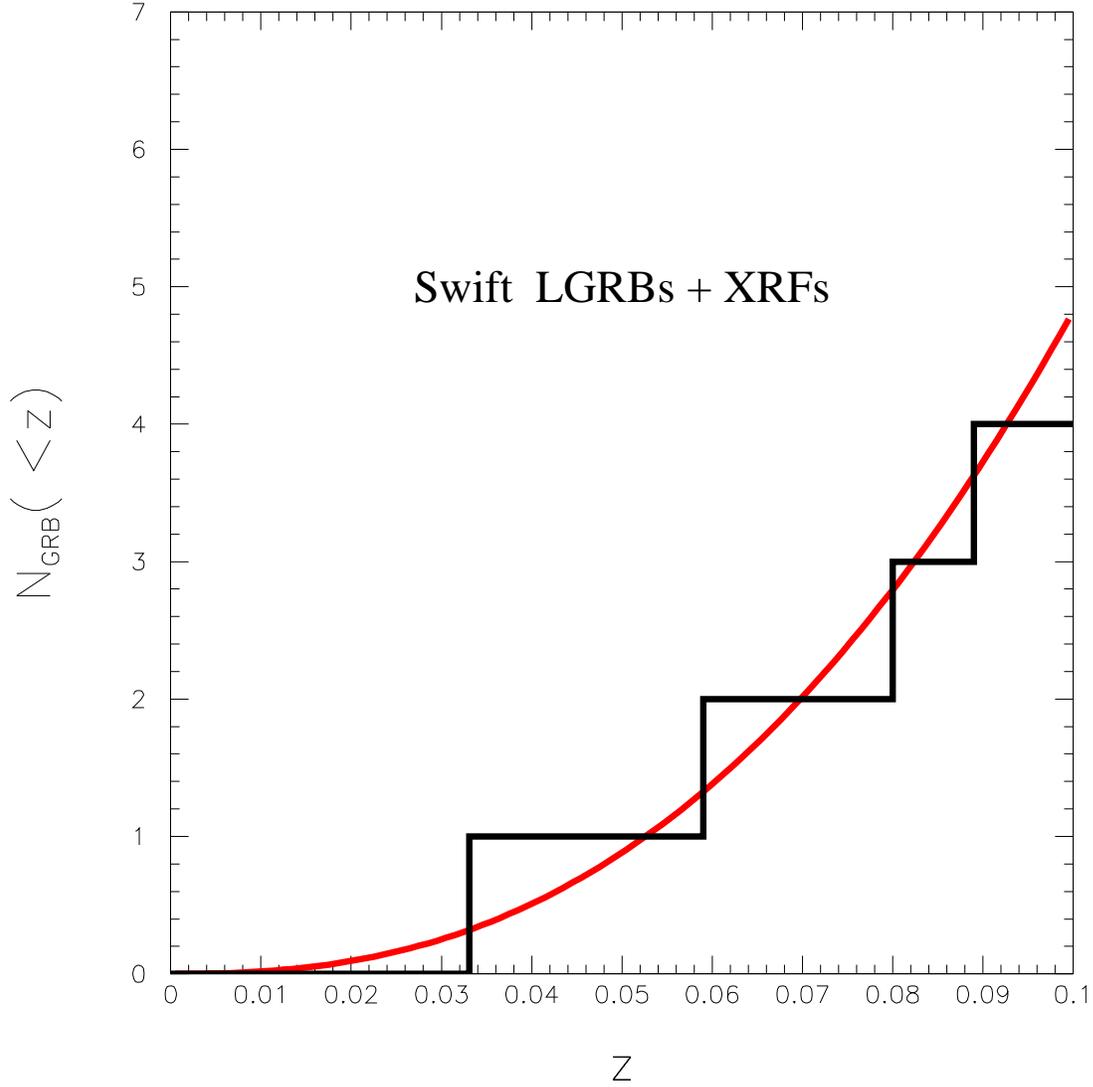,width=16.cm,height=16.cm}
\caption{Comparison between the cumulative distribution function $N(<z)$
of LGRBs with redshift below 0.1 (hystogram)  
that were detected by Swift before November 15, 2013 
and the CB model prediction 
of $N(<z)$ for $z<0.1$ as given by Eq.~(8). } 
\label{Fig9}
\end{figure}
\end{document}